\documentstyle[12pt]{article}

\newcommand{\be}{\begin{equation}}
\newcommand{\ee}{\end{equation}}

\def\psnormal{\textwidth=16cm\textheight=21.5cm
          \oddsidemargin=0.5cm\evensidemargin=0cm
          \topmargin=0cm\parindent=1cm}
\psnormal

\begin{document}
\pagestyle{empty}

\hspace{3cm}

\vspace{-3.4cm}
\rightline{{ CERN--TH.6764/92}}
\rightline{{ IEM--FT--66/92}}
\rightline{{ FTUAM 92/45}}

\vspace{0.4cm}
\begin{center}
{\bf A {\large N}ATURAL {\large S}OLUTION
TO THE {\large $\mu$} {\large P}ROBLEM}
\vspace{0.8cm}

J.A. CASAS${}^{*,**}\;$ and C. MU\~NOZ${}^{***}$
\vspace{0.8cm}

${}^{*}$ CERN, CH--1211 Geneva 23, Switzerland
\vspace{0.4cm}

${}^{**}$ Instituto de Estructura de la Materia (CSIC),\\
Serrano 123,  E-28006 Madrid, Spain
\vspace{0.4cm}

${}^{***}$ Dept. de F\'{\i}sica Te\'orica C-XI, \\
Univ. Aut\'onoma de Madrid, E-28049 Madrid, Spain
\vspace{0.4cm}

\end{center}

\centerline{\bf Abstract}
\vspace{0.3cm}

\noindent
We propose a simple mechanism for solving the $\mu$ problem
in the context of minimal low--energy
supergravity models. This is based on the appearance
of non--renormalizable couplings in the superpotential. In particular,
if $H_1H_2$ is an allowed operator by all the symmetries of the theory,
it is natural to promote the usual renormalizable superpotential $W_o$
to $W_o+\lambda W_o H_1H_2$, yielding an effective $\mu$ parameter
whose size is directly related to the gravitino mass once supersymmetry
is broken (this result
is maintained if $H_1H_2$ couples with different strengths
to the various terms present in $W_o$).
On the other hand, the $\mu$ term must be absent from $W_o$, otherwise
the natural scale for $\mu$ would be $M_P$.
Remarkably enough, this is entirely
justified in the supergravity theories
coming from superstrings, where mass terms for light fields are
forbidden in the superpotential.
We also analyse
the $SU(2)\times U(1)$ breaking, finding that it takes place
satisfactorily.
Finally, we give a realistic example in which supersymmetry
is broken by gaugino condensation,
where the mechanism proposed for solving the $\mu$ problem
can be gracefully implemented.

\vspace{0.3cm}
\begin{flushleft}
{CERN--TH.6764/92} \\
{IEM--FT--66/92} \\
{FTUAM 92/45} \\
{December 1992}
\end{flushleft}
\psnormal
\newpage
\mbox{}

\newpage
\pagestyle{plain}
\pagenumbering{arabic}
\section{Introduction}

One of the interesting features of low--energy supergravity (SUGRA)
models is that the electroweak symmetry breaking can be a direct
consequence of supersymmetry (SUSY) breaking [1]. In the ordinary
SUGRA models, SUSY breaking takes place in
a hidden sector of the theory, so that the gravitino mass $m_{3/2}$
becomes of the electroweak scale order. Below the Planck mass,
$M_P$, one is left with a global SUSY Lagrangian plus some terms
(characterized by the $m_{3/2}$ scale) breaking explicitly, but softly,
global SUSY. As we will briefly review below, the breakdown of
$SU(2)\times U(1)_Y$ appears as an automatic consequence of the
radiative corrections to these terms. The so--called $\mu$ problem
[2] arises in this context.

Let us consider a SUGRA theory with superpotential $W(\phi_i)$
and canonical kinetic terms  for the $\phi_i$ fields\footnote{We will
consider this case throughout the paper for simplicity. Our general
conclusions will not be modified by taking a more general case.}.
Then, the scalar potential takes the form [3]
\begin{eqnarray}
V = e^K \left[ \sum_i \left| \frac{\partial W}{\partial \phi_i}
+ \bar{\phi_i}W \right|^2 - 3|W|^2 \right]\;+\;\mathrm{D}\;\mathrm{terms}
\;\;,
\label{V}
\end{eqnarray}
where $K=\sum_i|\phi_i|^2$ is the K\"ahler potential. It is customary
to consider $W$ as a sum of two terms corresponding to the observable
sector $W^{obs}(\phi_i^{obs})$ and a hidden sector
$W^{hid}(\phi_i^{hid})$
\begin{eqnarray}
W(\phi_i^{obs},\phi_i^{hid})=
W^{obs}(\phi_i^{obs}) + W^{hid}(\phi_i^{hid})
\;\;.
\label{W}
\end{eqnarray}
$W^{hid}(\phi_i^{hid})$ is assumed to be
responsible for the SUSY breaking,
which implies that some of the $\phi_i^{hid}$ fields
acquire non--vanishing vacuum expectation values (VEVs) in the process.
Then, the form of the effective
observable scalar potential obtained from eq.(\ref{V}), assuming
vanishing cosmological constant, is [4]
\begin{eqnarray}
V^{obs}_{eff} = \sum_i \left| \frac{\partial \hat{W}^{obs}}
{\partial \phi_i^{obs}}\right|^2 &+& m_{3/2}^2 \sum_i | \phi_i^{obs}|^2
+ \left( Am_{3/2}\hat W^{obs}_t + Bm_{3/2}\hat W^{obs}_b\; +\;
\mathrm{h.c.} \right)
\nonumber \\
\;&+&\;\mathrm{D}\;\mathrm{terms}
\label{Vobs}
\end{eqnarray}
with
\begin{eqnarray}
m_{3/2}^2 = e^{K^{hid}}|W^{hid}|^2
\label{m32}
\end{eqnarray}
\begin{eqnarray}
B = A-1=\sum_i \left( | \phi_i^{hid}|^2 + \frac{\bar{\phi}^{hid}}
{\bar{W}^{hid}}\frac{\partial \bar{W}^{hid}}
{\partial \bar{\phi}_i^{hid}}\right)\;-\;1\;\;,
\label{AB}
\end{eqnarray}
where $K^{hid}=\sum_i |\phi_i^{hid}|^2$, $\hat W^{obs}$ is the
rescaled observable superpotential $\hat W^{obs}=e^{K^{hid}/2}
W^{obs}$, the subindex $t$($b$) denotes the trilinear (bilinear) part of
the superpotential, and $A$, $B$ are dimensionless numbers of $O(1)$,
which depend on the VEVs of the hidden fields. Since we are assuming
that SUSY breaking takes place at a right scale, the gravitino mass
given by eq.(\ref{m32}) is hierarchically smaller than the Planck
mass (i.e. of order the electroweak scale).

In the minimal supersymmetric standard model (MSSM) the matter content
consists of three generations of quark and lepton superfields plus
two Higgs doublets, $H_1$ and $H_2$, of opposite hypercharge.
Under these conditions the most general effective observable
superpotential has the form
\begin{eqnarray}
W^{obs}=\sum_{generations}(h_uQ_LH_2u_R +
h_dQ_LH_1d_R + h_eL_LH_1e_R )+ \mu H_1H_2\;\;.
\label{Wobs}
\end{eqnarray}
This includes the usual Yukawa couplings (in a self--explanatory
notation) plus a possible mass term for the Higgses, where $\mu$
is a free parameter. From eq.(\ref{Vobs}) the relevant Higgs
scalar potential along the neutral direction for the electroweak
breaking is readily obtained
\begin{eqnarray}
V(H_1,H_2)=\frac{1}{8}(g^2+g'^2)\left(|H_1|^2-|H_2|^2\right)^2
+ \mu_1^2|H_1|^2 + \mu_2^2|H_2|^2 -\mu_3^2(H_1H_2+\mathrm{h.c.})
\;\;,
\label{Vhiggs}
\end{eqnarray}
where
\begin{eqnarray}
\mu_{1,2}^2 &=& m_{3/2}^2 + \hat\mu^2
\nonumber \\
\mu_{3}^2 &=& -Bm_{3/2}\hat\mu
\nonumber \\
\hat\mu &\equiv & e^{K^{hid}/2}\mu\;\;.
\label{mus}
\end{eqnarray}
This is the SUSY version of the usual Higgs potential in the standard
model. In order for the potential to be bounded from below,
the condition
\begin{eqnarray}
\mu_{1}^2+\mu_2^2-2|\mu_3^2|>0
\label{condmus}
\end{eqnarray}
must be
imposed all over the energy range $[M_Z,M_P]$. This implies in particular
$\langle H_{1,2}\rangle = 0$ at the Planck scale. Below the Planck
scale, one
has to consider the radiative corrections to the scalar potential.
Then the boundary conditions of eq.(\ref{mus}) are substantially
modified in such a way that the determinant of the Higgs
mass--squared matrix
becomes negative, triggering $\langle H_{1,2}\rangle \neq 0$
and $SU(2)\times U(1)_Y$ symmetry breaking [1].

For this scheme to work, the presence of the last term in
eq.(\ref{Wobs}) is crucial. If $\mu=0$, then the form of the
renormalization group equations (RGEs) implies that such a term
is not generated at any $Q$ scale since $\mu(Q)\propto \mu$.
The same occurs for $\mu_3$, i.e. $\mu_3(Q)\propto \mu$.
Then, the minimum of the potential of eq.(\ref{Vhiggs}) occurs
for $H_1=0$ and, therefore, $d$--type quarks
and $e$--type leptons remain massless. Besides, the superpotential
of eq.(\ref{Wobs}) with $\mu=0$ possesses a spontaneously broken
Peccei--Quinn
symmetry [5] leading to the appearance of an unacceptable
Weinberg--Wilczek axion [6].

Once it is accepted that the presence of the $\mu$ term
in the superpotential is essential, there arises an inmediate
question: Is there any dynamical reason why $\mu$ should be
small, of the order of the electroweak scale? Note that,
to this respect, the $\mu$ term is different from the SUSY
soft--breaking terms,
which are characterized by the small scale $m_{3/2}$ once
we assume correct SUSY breaking. In principle the natural scale
of $\mu$ would be $M_P$, but this would reintroduce the
hierarchy problem since the Higgs scalars get a contribution
$\mu^2$ to their squared mass [see eq.(\ref{mus})]. Thus,
any complete explanation of the electroweak breaking scale
must justify the origin of $\mu$. This is the so--called
$\mu$ problem [2]. This problem has been considered by
several authors and different possible solutions
have been proposed [2,7,8].
In this letter we suggest a scenario in which $\mu$ is
generated by non--renormalizable terms
and its size is directly related to the gravitino mass.
A comparison with the scenarios of refs.[2,7,8] is also made.

\section{A natural solution to the $\mu$ problem}

Let us start with a simple scenario with superpotential
\begin{eqnarray}
W=W_o + \lambda W_oH_1H_2
\;\;.
\label{WWo}
\end{eqnarray}
where $W_o$ is the usual superpotential (including both observable
and hidden sectors) {\em without} a $\mu H_1H_2$ term. We have allowed
in (\ref{WWo}) a non--renormalizable term, characterized by the coupling
$\lambda=O(1)$ (in Planck units), which mixes the observable sector with
the hidden sector (other higher--order terms of this kind could also be
included, but they are not relevant for the present analysis).
The $\mu H_1H_2$ term must be absent from $W_o$ since, as was
mentioned above, the natural scale for $\mu$ would otherwise be $M_P$.
Certainly, this is technically possible in a supersymmetric theory,
since the non--renormalization theorems assure that this term cannot
be generated radiatively if initially $\mu=0$. One may wonder, however,
whether there is a theoretical reason for the absence of the
$\mu H_1H_2$ term from $W_o$ in eq.(\ref{WWo}), since it is not forbidden by
any
symmetry of the theory\footnote{The $\mu H_1H_2$ term can be
forbidden by invoking a Peccei--Quinn (PQ) symmetry [2,8]. This
is not possible here since (\ref{WWo}) does not possess any PQ
symmetry.}. It is quite remarkable here that this is provided
in the low--energy SUSY theory obtained
from superstrings. In this case mass terms (like $\mu H_1H_2$) are
forbidden in the superpotential. We will see in section 4 an explicit
example in this context. Finally, non-renormalizable terms
(like $\lambda W_oH_1H_2$) are in principle allowed in a generic
SUGRA theory.
Next, we show that
the $\lambda W_oH_1H_2$ term yields dynamically a $\mu$ parameter.

Using the general expression of eq.(\ref{V}), the scalar
potential $V$ generated by $W$ has the form
\begin{eqnarray}
V = &e^K& \left\{ \sum_i \left| \frac{\partial [W_o(1+\lambda
H_1H_2)]}{\partial \phi_i}
+ \bar{\phi_i}W_o (1+\lambda H_1H_2)\right|^2 - 3|W_o(1+\lambda
H_1H_2)|^2 \right\}
\nonumber\\
\;&+&\;\mathrm{D}\;\mathrm{terms}
\;\;,
\label{V2}
\end{eqnarray}
which can be written as
\begin{eqnarray}
V = V^{(1)}|1+\lambda H_1H_2|^2 &+&
e^K \left\{ \left| \frac{\partial [W_o(1+\lambda
H_1H_2)]}{\partial H_1}
+ \bar{H_1}W_o (1+\lambda H_1H_2)\right|^2 +(H_1\leftrightarrow H_2)
\right\}
\nonumber\\
\;&+&\;\mathrm{D}\;\mathrm{terms}
\;\;,
\label{V3}
\end{eqnarray}
where
\begin{eqnarray}
V^{(1)}\equiv e^K \left( \sum_i \left| \frac{\partial W_o}
{\partial \phi_i}
+ \bar{\phi_i}W_o \right|^2 - 3|W_o|^2 \right)\;
;\;\;\phi_i\neq H_{1,2}\;\;.
\label{V1}
\end{eqnarray}
Since $H_{1,2}$ enter in $W_o$ only through the ordinary Yukawa
couplings and we are assuming vanishing VEVs for the observable
scalar fields, it is clear  (recall that $W_o$ does not contain
a $\mu H_1H_2$ coupling) that $\left. \frac{\partial W_o}
{\partial H_{1,2}}\right|_{min}=0$. Besides, the vanishing
of the cosmological constant implies $V^{(1)}=0$ at the minimum
of the potential.
So, we can extract from the second term in eq.(\ref{V3}) the soft
terms associated with $H_{1,2}$:
\begin{eqnarray}
V(H_1,H_2)=&\frac{1}{8}&(g^2+g'^2)\left(|H_1|^2-|H_2|^2\right)^2
+ m_{3/2}^2(1+\lambda^2)|H_1|^2 + m_{3/2}^2(1+\lambda^2)|H_2|^2
\nonumber\\
&+&2m_{3/2}^2\lambda(H_1H_2+\mathrm{h.c.})
\;\;.
\label{Vhiggs2}
\end{eqnarray}
Comparing eqs.(\ref{Wobs}--\ref{mus}) with
eqs.(\ref{WWo},\ref{Vhiggs2}) it is clear that $\lambda W_oH_1H_2$
behaves like a $\mu$ term when $W_o$ acquires a non--vanishing VEV
dynamically. Defining $\lambda\langle W_o\rangle\equiv\mu$ we
can write eq.(\ref{Vhiggs2}) as eqs.(\ref{Vhiggs},\ref{mus}) where
now the value of $B$ is
\begin{eqnarray}
B=2
\;\;.
\label{B}
\end{eqnarray}
The value of $A$ is still given by eq.(\ref{AB}), but the relation
$B=A-1$ is no longer true. The fact that the new
"$\mu$ parameter" is of the electroweak--scale order
is a consequence of our assumption of a correct SUSY--breaking
scale $m_{3/2}=e^{K/2}W=O(M_Z)$. Finally, note that the usual
condition for the potential to be bounded from below
(\ref{condmus}) is automatically satisfied by (\ref{Vhiggs2})
for any value of $\lambda$.

One may wonder how general is the simple scenario of eq.(\ref{WWo}).
First of all, let us note that the fact that $H_1H_2$ is not forbidden
by any symmetry of the theory is a key ingredient for this scenario to
work. An obvious generalization of (\ref{WWo}) arises when $W_o$
consists of several
terms $W_o=W_o^{(1)}+W_o^{(2)}+...$ and $H_1H_2$ couples with
a different strength to each term, i.e. $(\lambda_1W_o^{(1)}+
\lambda_2W_o^{(2)}+...)H_1H_2$. However, provided that the hierarchical
small value for $\langle W_o\rangle$ is not achieved by a fine--tuning
between the VEVs of the various terms $W_o^{(1)},W_o^{(2)},...$,
it is clear that the order of magnitude of $\mu$ continues being
$m_{3/2}$. Apart from this, it should be noticed that $\lambda_i=O(1)$
(in Planck units) is only natural if $W_o^{(i)}$ is not an operator
with a extremely small coupling constant. However, this would be a
naturalness problem by itself. This would happen, for instance,
for $W_o^{(i)}=m\Phi^2$ with $m<<M_P$. (These terms
are forbidden in string theories.)

%
To conclude this section, it is worth noticing that in the context of
supergravity theories there is another
possible solution to the $\mu$ problem. Since the K\"ahler potential
$K$ is an arbitrary real--analytic function of the scalar fields,
we can study for example a theory with the following
$K$
\begin{eqnarray}
K=\sum_i|\phi_i|^2 + f(g(\phi_j, \bar \phi_j)H_1H_2\ +\ \mathrm{h.c.})
\;\;,
\label{K}
\end{eqnarray}
where $\phi_j\neq H_{1,2}$ and $f$ and $g$ are generic functions
($\langle g(\phi_j, \bar\phi_j)\rangle= O(1)$). Then, although $W_o$
does not contain a $\mu$ term, this is generated in the
scalar potential. This is trivial to see for the simplest case
(i.e. $f(x)=x$, $g=$ const. $\equiv\lambda$). Then the theory is equivalent
to one with K\"ahler potential $\sum_i|\phi_i|^2$ and superpotential
$W_oe^{\lambda H_1H_2}$, since the function ${\cal G}=K+\log|W|^2$
that defines the SUGRA theory is the same for both. Expanding
the exponential, the first two terms coincide with eq.(\ref{WWo})
and hence we obtain the same $\mu$ term as in eq.(\ref{Vhiggs2}).
The possibility (\ref{K}) was examined in ref.[7] for $f(x)=x$ and
when $g$ is a non--trivial function of the hidden fields,
in particular for the simplest case $g(\phi_j, \bar\phi_j)=\bar\xi$,
where $\bar\xi$ is a hidden field. It remains to be explored whether
a K\"ahler potential similar to that of eq.(\ref{K}) can arise in
the context of superstring theories.

\section{Expectation values for the Higgses}

In the above analysed solution to the $\mu$ problem it is assumed
that the observable scalar fields have vanishing VEVs at the Planck
scale. Since the non--renormalizable term $\lambda W_oH_1H_2$
mixes observable and hidden fields, one may wonder whether that
assumption is still true for the Higgses. We will show now that
this is in fact the case.

We assume here that the initial superpotential $W_o$
gives a correct SUSY breaking,
i.e. small gravitino mass and vanishing cosmological
constant. This means that $V_o$, i.e. the scalar potential
derived from $W_o$, is vanishing at the minimum
$\left. V_{o}\right|_{min}=0$ and thus
positive--definite. Using
the general expression of eq.(\ref{V}), $V_o$
can be decomposed in three pieces
\begin{eqnarray}
V_o = V^{(1)}\;+\;e^K \left\{ \left| \frac{\partial W_o}{\partial H_1}
+ \bar{H_1}W_o \right|^2 + (H_1\rightarrow H_2)\right\}\;+\;
\mathrm{D}\;\mathrm{terms}
\;\;,
\label{Vo}
\end{eqnarray}
where $V^{(1)}$ is defined in eq.(\ref{V1}). Recalling that we
are assuming that $W_o$ does not contain a $\mu H_1H_2$ term
and that $\left. \frac{\partial W_o}{\partial H_{1,2}}\right|_{min}=0$
(since squarks and sleptons are supposed to have vanishing VEVs),
it is clear that $V^{(1)}$ is flat in $H_{1,2}$.
So, the minimum of the second piece of (\ref{Vo}) is zero and occurs
at $H_{1,2}=0$ (for any value of $W_o$). Therefore, necessarily
$\left. V^{(1)}\right|_{min}=0$, i.e. $V^{(1)}$ is also
positive--definite. All this is very ordinary: it simply means that
the hidden sector is entirely responsible for the breaking.
(Note that the $H_{1,2}$ F--terms are vanishing, while some
of the hidden fields F--terms must be different from zero.)
Notice also that from (\ref{Vo}) one obtains
$e^{K}|W_o|^2(|H_1|^2+|H_2|^2)=
m_{3/2}^2|H_1|^2+m_{3/2}^2|H_2|^2$
but, because of the absence of a $\mu H_1H_2$ term in $W_o$, there
is no $Bm_{3/2}\hat\mu H_1H_2$ term in the scalar potential.

Let us now study the impact
of doing, according to our approach, $W_o \rightarrow W = W_o + \lambda
W_oH_1H_2$. The corresponding scalar
%
%
%
potential, $V$, has already been written in eq.(\ref{V3}). Now, since
$V^{(1)}$ is positive--definite, so is $V$. In fact, the minimum of $V$
is for $V=0$ and occurs when the three pieces of (\ref{V3}) are
vanishing. Clearly, the minimum of the first and third pieces
of (\ref{V3}) coincides with that of eq.(\ref{Vo}) above, implying
$\left. V^{(1)}\right|_{min}=0$,\footnote{The only exception occurs if
$\lambda H_1H_2=-1$,
but then the second piece of (\ref{V3}), which is also
positive--definite, is different from zero, so this is not a
solution for the minimization of the whole potential.}
and thus the VEV of $W_o$ is the
same
as when we started with just $W_o$. Finally,
recalling that
$\left. \frac{\partial W_o}{\partial H_{1,2}}
\right|_{min}=0$, it is clear that the second piece of $V$
in eq.(\ref{V3}) has two possible minima
%
%
\begin{eqnarray}
H_1, H_2=0
\;\;,
\label{Hs1}
\end{eqnarray}
\begin{eqnarray}
\lambda H_2+(1+\lambda H_1H_2)\bar H_1&=&0
\nonumber\\
(H_1\leftrightarrow H_2)&=&0
\label{Hs2}
\end{eqnarray}
As was explained in section 1, the solution (\ref{Hs1}) is the
phenomenologically interesting one, whereas
the solution (\ref{Hs2}) leads to $H_{1,2}\sim M_P$,
so it is not phenomenologically viable. We can ignore this
solution since if $H_{1,2}$ are initially located at $H_{1,2}=0$
(e.g. by thermal effects) they will remain there as long as (\ref{Hs1})
continues to be a minimum solution. Of course, radiative corrections
will trigger non--zero VEVs of the correct size for $H_1$, $H_2$.

\section{A realistic example}

As we saw in section 2, the assumption of correct
SUSY breaking was crucial for obtaining the $\mu$ parameter
of the electroweak--scale order. As a matter of fact, gaugino
condensation effects in the hidden sector [9] are the most
satisfactory mechanism so far explored, able to break SUSY at
a scale hierarchically smaller than $M_P$ [10]. The reason
is that the scale of gaugino condensation corresponds to the
scale at which the gauge coupling becomes large, and this is
governed by the running of the coupling constant. Since
the running is only logarithmically dependent on the scale,
the gaugino condensation scale is suppressed relative to the
initial one by an exponentially small factor $\sim e^{-1/2\beta g^2}$
($\beta$ is the one--loop coefficient of the beta function
of the hidden sector gauge group $G$). This mechanism has been
intensively studied in the context of SUGRA theories coming from
superstrings [11,12], where
the gauge coupling is related to the VEV of the dilaton field
$S$ (more specifically  Re$S=g^{-2}$). Recall that
we have argued in section 2 that superstring theories are precisely
a natural context where the solution of the $\mu$ problem presented
here can be implemented, since
mass terms, such as $\mu H_1 H_2$, appearing in the superpotential
are automatically
forbidden in superstrings. Besides, non--renormalizable terms
like $\lambda W_oH_1H_2$ in eq.(\ref{WWo})
are in principle allowed and, in fact, they are usually present [13].

In the absence of hidden matter, the condensation process is
correctly described by a non--perturbative effective
superpotential
\begin{eqnarray}
W_o\propto e^{-3S/2\beta_o}
\;\;,
\label{Wcond}
\end{eqnarray}
with $\beta_o=3C(G)/16\pi^2$, where $C(G)$ is the Casimir operator
in the adjoint representation of $G$. It is difficult to imagine,
however, how the mechanism expounded in section 2 could be implemented
here. More precisely, it is not clear
that we could have something like
$W=W_o+\lambda W_oH_1H_2$, due to the effective character
of (\ref{Wcond}).

Fortunately, things are different in the presence of hidden matter,
which is precisely the most frequent case in string constructions
[13]. There is not at present a generally accepted formalism
describing the condensation in the presence of massless matter,
but the case of massive matter is well understood [14]. For example,
in the case of $G=SU(N)$ with $M(N+\bar N)$ "quark" representations
$Q_\alpha$, $\bar Q_\alpha$,
$\alpha=1,...,M$, with a mass term given by
\begin{eqnarray}
W_o^{pert}=-\sum_{\alpha,\beta}{\cal M}_{\alpha,\beta}Q_{\alpha}
\bar Q_\beta
\;\;,
\label{Wpert}
\end{eqnarray}
the complete condensation superpotential can be written as [12]
\begin{eqnarray}
W_o\propto [\mathrm{det}{\cal M}]^{\frac{1}{N}} e^{-3S/2\beta_o}
\;\;.
\label{Wcond2}
\end{eqnarray}
It should be noticed here that, strictly speaking, there are no mass
terms like (\ref{Wpert}) in the context of string theories.
However the matter fields usually have trilinear couplings which
play the role of mass terms with a dynamical mass given by the VEV
of another matter field. The simplest case occurs when there is
an $SU(N)$ singlet field $A$ giving mass to all the quark representations.
Then (\ref{Wpert}) takes the form
\begin{eqnarray}
W_o^{pert}=-\sum_{\alpha=1}^M AQ_{\alpha}\bar Q_\alpha
\;\;,
\label{Wpert2}
\end{eqnarray}
and $\mathrm{det}{\cal M}=A^M$. Now, if $H_1H_2$ is an allowed
coupling from
all the symmetries of the theory, it is natural to promote
$W_o^{pert}$ to\footnote{We neglect here higher--order
non--renormalizable couplings since they do not contribute to the
$\mu$ term.}
\begin{eqnarray}
W^{pert}=-\sum_{\alpha}A(1+\lambda' H_1H_2)Q_{\alpha}\bar Q_\alpha
\;\;,
\label{Wpert3}
\end{eqnarray}
so that $\mathrm{det}{\cal M}=[A(1+\lambda' H_1H_2)]^M$, and
(\ref{Wcond2}) takes the form
\begin{eqnarray}
W_o\rightarrow W\propto
[A(1+\lambda' H_1H_2)]^{\frac{M}{N}} e^{-3S/2\beta_o}
\simeq A^{\frac{M}{N}}(1+\frac{M}{N}\lambda' H_1H_2) e^{-3S/2\beta_o}
\;\;.
\label{Wcond3}
\end{eqnarray}
Thus
\begin{eqnarray}
W=W_o+\lambda W_oH_1H_2
\;\;,
\label{WWocond2}
\end{eqnarray}
where we have defined $\lambda\equiv\frac{M}{N}\lambda'$. This is
precisely the kind of superpotential we wanted (see eq.(\ref{WWo}))
in order to generate the $\mu$ term dynamically.

In ref.[8] an interesting solution to the $\mu$ problem was proposed
in a similar context with a PQ symmetry, using the presence of a term
$H_1H_2Q\bar Q$
in the superpotential and assuming that the scalar components
of $Q$ and $\bar Q$ condense at a scale $\Lambda\simeq 10^{11}$ GeV.
As mentioned above, the only accepted formalism
describing the
condensation is in the presence of massive matter. Thus the previous
term
behaves as a dynamical mass term for the squarks
and the complete superpotential (\ref{Wcond2}) becomes
$W\propto (H_1H_2)^{\frac{1}{N}}
e^{-3S/2\beta_o}$. This is phenomenologically unviable since
the Higgses must have vanishing VEVs at $M_P$ for a correct
phenomenology, which would imply $\langle W\rangle=0$ and thus
no SUSY breaking. We can improve this model by including
a mass term for $Q\bar Q$. However, a genuine mass term for $Q\bar Q$
would break
the PQ symmetry, so one should consider something similar to
(\ref{Wpert2}). Then the perturbative superpotential is
\begin{eqnarray}
W^{pert}\sim  AQ\bar Q + H_1H_2 Q\bar Q
\;\;,
\label{Wpert4}
\end{eqnarray}
and the scenario becomes much more similar to that given
by eq.(\ref{Wpert3}). However, there  still is an important
difference. In eq.(\ref{Wpert3}) $H_1H_2$ couples to
$AQ\bar Q$ (which is the natural thing if $H_1H_2$ is invariant
under all the symmetries of the theory) instead of $Q\bar Q$;
thus there is no PQ symmetry. Moreover, (\ref{Wpert3}) leads
to (\ref{WWocond2}) in which the $\mu$ scale is directly given
by the $m_{3/2}$ scale ($\mu=O(m_{3/2})$). However from
(\ref{Wpert4}) the $\mu$ scale is given by the
squark condensation scale [12] $\langle Q\bar Q\rangle/M_P\simeq
m_{3/2}M_P/N\langle A\rangle$, so that the value of $\mu$ in this case
tends to be a bit too large.

\section{Summary and conclusions}

We have proposed a simple mechanism for solving the $\mu$ problem
in the context of minimal low--energy SUGRA models. This is based
on the appearance
of non--renormalizable couplings in the superpotential. In particular,
if $H_1H_2$ is an allowed operator by all the symmetries of the theory,
it is natural to promote the usual renormalizable superpotential $W_o$
to $W_o+\lambda W_o H_1H_2$, yielding an effective $\mu$ parameter
whose size is directly related to the gravitino mass once SUSY
is broken (this result
is essentially maintained if $H_1H_2$ couples with different strengths
to the various terms present in $W_o$).

On the other hand, the $\mu$ term must be absent in $W_o$, otherwise
the natural scale for $\mu$ would be $M_P$.
Certainly this is technically possible in a supersymmetric theory
since the non--renormalization theorems assure that
this term cannot be generated radiatively if initially $\mu=0$. Remarkably
enough, however,
a theoretical reason for the absence of the
$\mu H_1H_2$ term from $W_o$ is provided
in the low--energy SUSY theory obtained
from superstrings. In this case mass terms (such as $\mu H_1H_2$) are
forbidden in the superpotential (however, non--renormalizable terms
like $\lambda W_oH_1H_2$
are in principle allowed and, in fact, they are usually present).

We have also addressed other alternative solutions, comparing them
with the one proposed here. On the other hand, we have analysed
the $SU(2)\times U(1)$ breaking, finding that it takes place
satisfactorily.

Finally, we have given a realistic example in which SUSY
is broken by gaugino condensation in the
presence of hidden matter (which is the usual situation in strings),
and where the mechanism proposed for solving the $\mu$ problem
can be gracefully implemented.


\vspace{2cm}
\noindent{\bf ACKNOWLEDGEMENTS}

We gratefully acknowledge J. Louis for extremely useful discussions.



\end{document}